# AN INTEGRATED CIRCUIT COMPATIBLE COMPACT PACKAGE FOR THERMAL GAS FLOWMETERS

*P. Bruschi, V. Nurra*

Dipartimento di Ingegneria dell'Informazione, via G. Caruso 16, I-56122, Pisa, Italy

*M. Piotto*

IEIIT Pisa - CNR, via G. Caruso 16, I-56122, Pisa, Italy

**ABSTRACT**

An original packaging method suitable for integrated thermal mass flow sensors is presented. The method consists in the application of a plastic transparent adapter to the chip surface. The adapter is sealed to the chip surface by means of a thermal procedure. By this approach it is possible to selectively convey the fluid flow to reduced chip areas, avoiding contact with the pads. Fabrication and testing of a very compact flow sensor is described.

## 1. INTRODUCTION

The extreme miniaturization of integrated flow sensors results in a series of desirable characteristics, such as reduced power consumption, fast response times and high sensitivity. Nevertheless, miniaturization introduces also fragility, low resistance to atmospheric agents and connection difficulty. These drawbacks seriously hinder the diffusion of integrated flowmeters in many application fields, where expensive and bulky traditional devices are often still preferred. Then, the design of a proper package is essential to allow a silicon chip, carrying micrometric sensing structures, to handle gas or liquid flows without being damaged [1-3].

Several strategies have been proposed to accomplish this task. The simplest solution is that of placing the whole chip, including bonding wires, inside a channel where the fluid is made to flow [4-7]. Clearly, this is not applicable to liquid flows, due to the exposed pads and bonding wires. In addition, it should be noted that, in the case of thermal flowmeters, which are actually velocity sensors, it is often requested that the gas flow is accelerated in proximity of the sensing structure, in order to detect very low flow rates. Including the whole chip inside the gas flow prevents the channel cross-section to be reduced below several $mm^2$.

An interesting approach that does not solve the problem of sensitivity optimization but provides a reliable protection of the chip from both gas and liquid flows, is that of separating the chip and the flow by means of a thermally conductive membrane [8,9]. Unfortunately, since the whole chip or large portions of the membrane should be heated for correct operation of this kind of sensors, an important degradation of the sensor speed and power consumption should be expected.

On the opposite side stand sensors based on microchannels etched either on the same silicon substrate as the sensing structures or on a silicon/glass cover, bonded to the main chip [10,11]. The fluid flows into narrow microchannels avoiding direct contact with the pads. This approach is applicable only to real microfluidic applications, since only very reduced channel cross-sections can be obtained. Furthermore, due to the high temperatures or high voltages involved in the bonding procedure, real compatibility with on-chip electronic components has still to be demonstrated. Another obstacle to the application of such a technique to standard integrated circuit is the difficulty to obtain sealed channels due to the uneven chip surface.

In this work we propose a new packaging method that can be applied to integrated thermal flow sensors made up of arrangements of thermally insulated micrometric heaters and temperature probes. The package is based on a Polymethyl-Methacrylate (PMMA) adapter that can be easily connected to standard pipes and conveys the fluid flow to the chip surface avoiding direct contact with pads and bonding wires. Combination of the proposed adapter with most standard packages for integrated circuits is possible. The result of tests performed in nitrogen demonstrates the effectiveness of the approach.

## 2. DEVICE DESCRIPTION

The sensing structure is a classical differential flowmeter [12] made up of a heater placed between an upstream and a downstream temperature probe. The heater is a 1 k$\Omega$ polysilicon resistor positioned over a dielectric membrane suspended by means of four 45 degrees arms. The temperature probes are two thermopiles including 7 $p^+$-poly/$n^+$-poly thermocouples with the hot contacts at the edge of a dielectric cantilever beam and the cold contacts on bulk silicon. The chip was designed with the BCD6 (Bipolar-CMOS-DMOS) process of STMicroelectronics.

Thermal insulation of the dielectric membranes from the substrate has been obtained by means of a post-processing procedure.

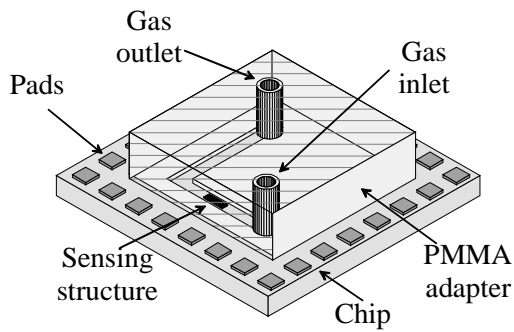

Fig. 1. Perspective view of the adapter - chip assembly.

The gas flow is conveyed to the sensing structure by means of an adapter, schematically shown in Fig. 1. A trench milled in the lower adapter face forms a channel sealed to the chip front face. The complete structure resulting from the chip-adapter connection is shown in Fig. 2 where the layout of the 4 x 4 mm$^2$ chip is also reported.. Details about the sensing structure fabrication and the procedure used to apply the adapter to the chip are reported in the two following sections.

**2.1. Post-processing**

The heater and the thermopile hot contacts were thermally insulated from the substrate by means of a cavity etched into the front-side of the chip. To this purpose, openings in the dielectric layers, through which the bare silicon substrate becomes accessible, must be provided. In some fabrication processes, like the BCD3s of STMicroelectronics [7,12] or other commercial CMOS processes [13], openings can be fabricated by the silicon foundry if a proper stacking of active area, contact, via and pad-opening layers is provided during the chip design. In this case, the post-processing is limited to the silicon etching.

Differently, in processes like that used in this work, the silicided active areas and tungsten plugs used to fill contacts and vias prevent a direct access to the bare silicon substrate. Since silicide and metal plug removal requires processes and equipments not available in a research laboratory, it is more convenient to perform both the dielectric openings and the silicon etching in the post-processing phase.

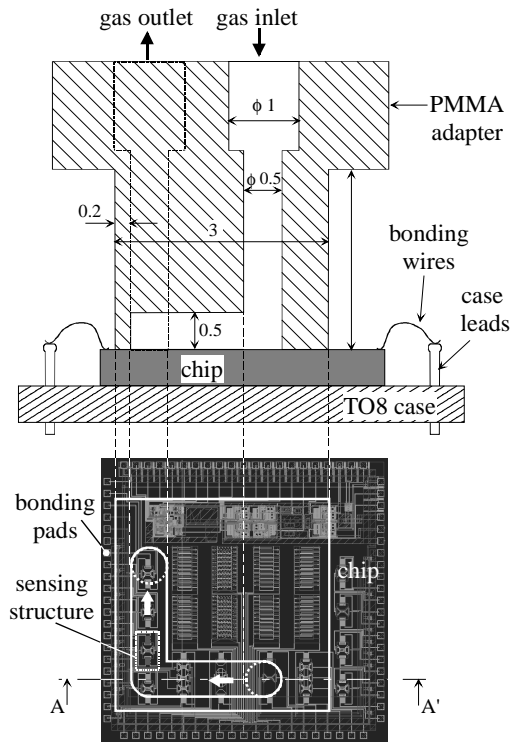

Fig. 2. Real structure of the adapter with the main dimensions (in mm) (not to scale).

In our case, the passivation openings have been performed by the silicon foundry while the residual dielectric layers have been removed in the post-processing. In figure 3 a photograph of the sensing structure before the post-processing is shown: the passivation openings are indicated.

The residual dielectric layers were removed by means a 1 μm resolution photolithography and a buffered HF(BHF) solution as $SiO_2$ etchant. An etching time of 30 minutes at room temperature was necessary to reach the bare silicon substrate. After that, silicon was anisotropically removed by means of a solution of 100 g of 5 wt% TMAH with 2 g of silicic acid and 0.6 g of ammonium persulfate. This modified TMAH solution has a good selectivity toward dielectric layers and aluminium allowing the silicon etch to be performed without any additional mask. Moreover, TMAH is less toxic than EDP solutions and is IC-compatible since it is used also as photoresist developer [14]. Nevertheless, the modified TMAH solutions are prone to a fast "aging-effect" with a noticeable decrease of the etch rate and increase of the etched surface roughness. A frequent solution refresh is thus necessary for long etch times [15]. In our case a 70 µm deep cavity has been obtained with only two etch steps, 40 minutes each, performed at 90 °C.

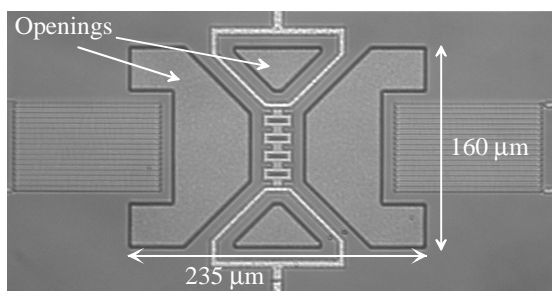

Fig. 3. Photograph of the sensing structure before post-processing.

The chip includes also a few electronic blocks that resulted fully functional after the post-processing technique described above. After the silicon etch, the die was glued to a 12 pin TO8 metal case by means of epoxy resin; wedge bonding was used to connect selected chip pads to the case pins.

Due to the small number of pins available in the package, only the devices strictly required for the flow-meter fabrication were connected. In particular the following blocks were involved: (i) a sensing structure, (ii) a ΔVbe substrate temperature sensor and (iii) a power mosfet. The latter, biased at fixed $V_{DS}$ and closed in feedback loop with the temperature sensor by means of an external op-amp, was used as a substrate heater to control the chip temperature. As it will be shown in detail in next section, this possibility has been exploited during the packaging phase

**2.2. Package design and assembling**

Packaging is perhaps the most critical phase in the flow sensor fabrication procedure. For integrated flow sensors, the package plays the important role of conveying the fluid flow to the sensing structures, avoiding at the same time contact between the fluid itself and non compatible chip elements such as the bonding pads.

The proposed packaging procedure is based on a adapter, schematically shown in Fig. 1, consisting in a solid piece with an optically flat face slightly smaller than the chip area free from bonding pads. A L-shaped trench is cut by means of a precision milling machine (VHF CAM 100) into the adapter face. The latter is aligned to the chip front side in order to include the sensing element into the trench and, at the same time, avoid contact with the bonding pads.

The real structure of the adapter with its main dimensions is shown in Fig. 2, together with the chip layout. Two holes connect the two extremes of the trench to the inlet and outlet fittings. The upper part of the adapter is enlarged with respect to the section joined to the chip in order to allow an easier connection to the gas line. Alignment of the adapter to the chip is obtained by means of a micrometric displacement stage, equipped with a low magnification microscope. During this phase, the adapter is kept separated from the chip by a small air gap (about 0.1 mm); then the adapter is placed in contact to the chip front side. Note that, using a purposely built holder, the adapter is left free to rotate along two axes parallel to the chip surface. In this way even a small angular mismatch is self-recovered applying a light pressure between the two pieces (chip and adapter). Unfortunately, this does not guarantee a leak free adhesion of the two surfaces, owing to the roughness of the chip front face, mainly due to the thick upper interconnect layers (metals) The latter, covered by non planarized dielectric passivation layers, results in protrusions acting as spacers for the two surfaces. Measurements performed by means of a stylus profilometer showed that, for the process adopted, steps up to 3 µm were present on the chip surface.

To overcome this problem and obtain good sealing of the trench (see fig. 1) we softened the adapter surface by heating the chip to a proper temperature. This, in combination of a light pressure applied between the two pieces, allowed the chip protrusion to penetrate into the adapter face sealing the trench. Heating of the chip during the application of the package was facilitated by the mentioned power mosfet and temperature sensor integrated on the same die as the sensing structures. Details about the circuit used to control the chip temperature are given at the end of this section.

In order to obtain the desired result, the choice of the material for the adapter is fundamental. The requisites are: (1) transparency, to facilitate the alignment of the sensing structure into the trench; (ii) acceptable surface hardness, to avoid unintentional scratches of the surface during alignment and (iii) a low enough glass transition temperature. We compared three commonly available, transparent plastic materials: PMMA, Polycarbonate and Polystyrene. The hardness and glass transition temperature are reported in table I for the three materials.

| Material | Hardness (Rockwell M) | Glass Transition temperature ( ºC) |
|---|---|---|
| PMMA | 90 | 100 |
| Polycarbonate | 70 | 150 |
| Polystyrene | 72 | 90 |

Table I. Properties of transparent polymers.

We choose PMMA for its higher hardness and relatively low glass transition temperature.

Sealing was obtained maintaining the chip for five minutes at 110 °C, a few degrees over the PMMA glass transition but still below its melting point (130 °C). A force of about 5 N was applied between the chip and the adapter during the softening phase. Finally, epoxy resin was carefully poured around the adapter perimeter to obtain a robust structure. It is important to observe that attempts to apply the resin without previous softening and sealing of the adapter surface resulted in penetration of the resin between the chip and adapter with eventual filling of the trench. The compactness of the final structure is shown in Fig. 4.

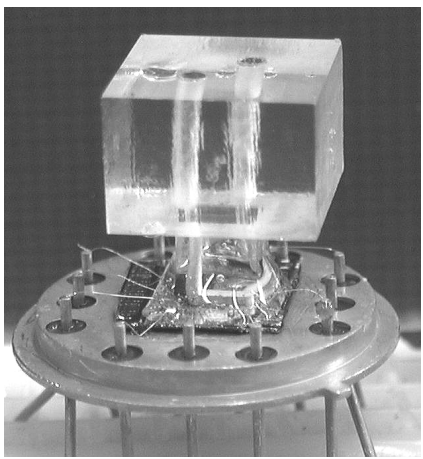

Fig. 4. Photograph of the final device.

The schematic diagram of the circuit used to control the chip temperature is shown in Fig. 5, where the on-chip components are included in the shaded box, while the other components are placed on a printed circuit board (PCB).

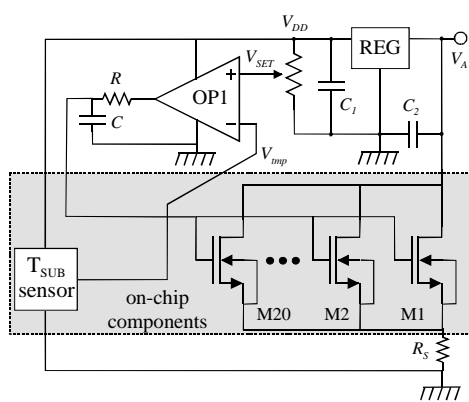

Fig. 5. Circuit used to control the chip temperature. The elements into the shaded box are integrated on the sensing chip.

The power mosfet is formed by 20 *n*-DMOS connected in parallel and distributed along the chip perimeter. In order to limit the maximum drain current, the resistor $R_S$ is placed in series with the source.

The circuit power supply voltage is $V_A$=5V. This constitutes the drain voltage of the power mosfet. A regulator is used to provide the 3.3 V power supply required by the chip. The $T_{SUB}$ sensor produces a voltage $V_{tmp}$ proportional to the absolute temperature of the silicon substrate with a sensitivity of 3 mV/K. An external operational amplifier, OP1, with rail-to-rail input and output ranges, compares the voltage $V_{tmp}$ with the set point voltage $V_{set}$ and drives the power mosfet gate. The R-C filter cuts high frequency components (>10 kHz) in order to avoid unwanted oscillations due to parasitic capacitive coupling.

## 3. EXPERIMENTAL RESULTS

The flowmeter characterization was performed at room temperature by connecting a nitrogen line to the inlet and outlet holes by means of stainless steel pipes of 1 mm external diameter. Sealing of the pipes to the adapter was obtained with silicone glue. The line, schematically shown in Fig. 6, was equipped with two reference flow controllers (MKS 1179B), one with a full scale range of 10 sccm and the other with a full scale range of 200 sccm, and a pressure gauge (Baratron® MKS 750B). The latter was connected upstream to the flowmeter under test in order to measure the insertion loss of the device.

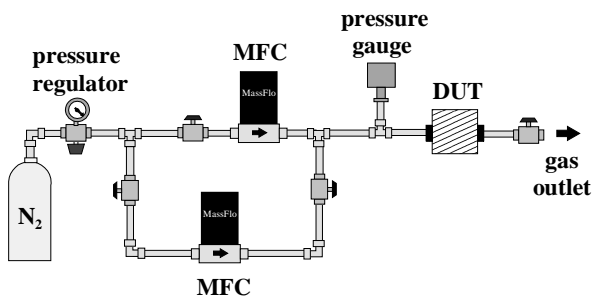

Fig. 6. Layout of the gas line used for device characterization: DUT indicates the device under test while MFC a mass flow controller.

A purposely built electronic circuit, mounted on a PCB, was used to supply the heater with a constant voltage and read the thermopile output signal. The schematic diagram is shown in Fig. 7. The input stage is based on the low noise instrumentation amplifier AD620 (IC1) set for a gain of 100 by means of resistor $R_G$. The two thermopiles of the sensing structure have one common terminal, connected to ground (input port $V_{CM}$).

The signal to be read is the difference between the remaining two thermopile terminals, connected to input ports $V_{T1}$ and $V_{T2}$.

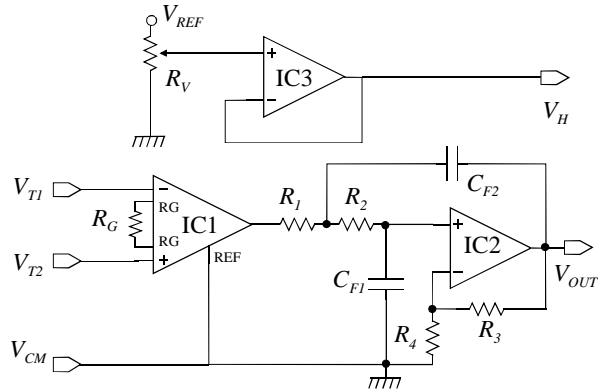

Fig 7. Circuit used to produce the heater voltage and read the output signal.

The heater resistor is driven by the operational amplifier IC3; the heater voltage can be adjusted through the variable resistor $R_V$. A 2nd order Butterworth low pass filter (IC2) with a cut off frequency of 10 Hz and gain 1.5 is used to reduce the noise bandwidth and improve resolution.

Figure 8 shows the dependence of the thermopile output voltage on the flow rate. The offset voltage present in zero flow condition, partly due to unavoidable asymmetries in the sensing microstructure, was electrically cancelled before starting the device characterization. Negative flows were measured by simply swapping the inlet and outlet connections.

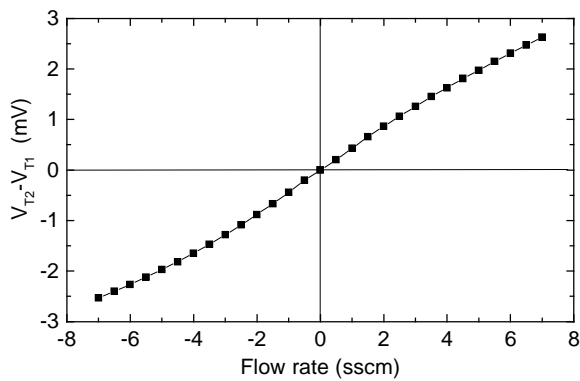

Fig. 8. Response of the sensor to nitrogen flow at room temperature.

A symmetrical response can be observed in the flow rate range of Fig. 4. The insertion loss was less than 100 Pa at the maximum flow rate value.

## 4. ACKNOWLEDGEMENTS

This work was financed by the Fondazione Cassa di Risparmio di Pisa (Italy) and by the company Laben, Florence. The authors thank the STMicroelectronics of Cornaredo (Italy) for fabricating the chips.